\documentclass[12pt]{article}
\usepackage{graphicx}
\begin{document}
\centerline{\Large\bf Phase Structure of Thermal QCD/QED:}
\centerline{\large\bf A Gauge Invariant Solution of }
\centerline{\large\bf the HTL Resummed Improved Ladder Dyson-Schwinger Equation}
\vspace*{0.5cm}
\centerline{Hisao NAKKAGAWA, Hiroshi YOKOTA and Koji YOSHIDA}
\centerline{Institute of Natural Sciences, Nara University, Nara 631-8502, Japan}
\vspace*{0.5cm}
{\bf Abstract}: Based on the hard-thermal-loop resummed improved ladder Dyson-Schwinger equation for
the fermion mass function, we propose a procedure how we can get the gauge invariant solution in the
sense it satisfies the Ward-Takahashi identity. Results of the numerical analysis are shown and
properties of the ``gauge-invariant'' solutions are discussed. 
\baselineskip=15pt
\vspace*{0.5cm}
\begin{flushleft}
{\large\bf 1. Introduction and Summary}
\end{flushleft}

The Dyson-Schwinger equation (DSE) has proven itself a powerful tool to investigate \textit{with the analytic
procedure} the nonperturbative structure of field theories, such as the chiral phase transition of gauge
theories. Actually analyses based on the DSE have successfully clarified the phase structure of
vacuum QEC/QCD [1,2,3]. Here we must take note of the fact that these DSE analyses in vacuum gauge
theories were carried out in the Landau gauge with the ladder interaction kernel. Same analyses have
been performed at finite temperature($T$) and/or density($\mu$) also in the Landau gauge with
the ladder interaction kernel [4,5,6], and with the hard-thermal-loop (HTL) resummed improved ladder
interaction kernel [7].

The reason why the ladder approximation is used is that the full DSEs are coupled integral equations
for several unknown functions, thus are hard to be solved without introducing appropriate approximations.
We usually adopt the step-by-step approach to this problem, firstly approximate the integration kernel by
the tree, or, ladder interaction kernel, next use the improved ladder one, etc. The possibility of such
a systematic improvement through the well-established analytic procedure is one of the important
characteristics of the DSE. In fact, results of Ref.[7] show that at finite temperature/density
it is important to correctly take the dominant thermal effect into the interaction kernel in terms
of the HTL resummation.

In case of the vacuum QED in the Landau gauge DSE with the ladder kernel for the fermion mass function,
the fermion wave function renormalization constant is guaranteed to be unity [1], namely the Ward-Takahashi
identity is automatically satisfied. Thus irrespective of the problem of the ladder approximation, the 
results obtained would be gauge invariant.  

At finite $T$ and/or $\mu$, however, there is no such guarantee. In fact, even in the Landau gauge
the fermion wave function renormalization constant largely deviates from unity [7,8], being not even real.
At finite $T$ and/or $\mu$ the results obtained from the ladder DSE explicitly violate the Ward-Takahashi
identity, thus depend on the choice of gauge. All the preceding analyses [4-7] suffer from the gauge-dependence
problem coming from the ladder approximation of the interaction kernel, their physical meaning being obscure. 

In this paper, we present, in the analysis of the HTL resummed improved ladder DS equation for the fermion 
mass function in thermal QED/QCD, the procedure to get the gauge invariant solution in the sense it satisfies 
the Ward-Takahashi identity. We firstly show that the solutions of the HTL resummed improved ladder DS equation 
in thermal QED/QCD suffer from the problem of gauge-parameter dependence, then solve numerically the DSE 
constrained to satisfy the Ward-Takahashi identity and investigate the properties of the ``gauge invariant'' 
solution. Part of the preliminary result of the analysis was reported in Ref.[9], showing the effectiveness 
of the procedure.

Results of the present analysis are summarized as follows: \\
    (1) The solution of the HTL resummed improved ladder DS equation depends strongly on the choice of the 
gauge parameter within the momentum-independent gauge. This type of solution always shows the explicit 
contradiction with the Ward-Takahashi identity. \\
    (2) We can determine numerically the solution that satisfies the Ward-Takahashi identity, namely 
the solution in which the fermion wave function renormalization constant is almost equal to unity. 
To get such a solution it is essential that we work in the nonlinear gauge where the gauge parameter $\xi$ 
depends on the  momentum of the gauge boson. \\
    (3) The chiral phase transition in the massless thermal QED/QCD is confirmed to occur through the 
second order transition; a dynamical fermion mass is generated at the critical temperature or at the 
critical coupling constant without discontinuity. \\
    (4) The effect of thermal fluctuation on the chiral symmetry breaking and/or restoration is smaller 
than that expected in the previous analysis in the Landau gauge [7].

\vspace*{0.5cm}
\begin{flushleft}
{\large\bf 2. DS equation for the fermion self-energy function $\Sigma_R$} \\
\vspace*{0.4cm}
{\bf 2-1. DS equation in the HTL resummed improved ladder approximation}
\end{flushleft}

 The fermion self-energy function $\Sigma_R$ appearing in the fermion propagator $S_R$
\begin{equation}
 S_R(P)  = [ P\!\!\!\!/ + i \epsilon \gamma^0 - \Sigma_R(P)]^{-1} 
\end{equation}
can be decomposed at finite temperature and/or density as 
\begin{equation}
    \Sigma_R(P) =  (1 - A(P)) p_i \gamma^i - B(P) \gamma^0   + C(P)
\end{equation}
with $A(P)$, $B(P)$ and $C(P)$ being the three scalar invariants to be determined.  In the present analysis, 
we use the HTL resummed form $^*G_{\mu \nu}$ for the gauge boson propagator,
\begin{equation}
{}^*G^{\mu\nu} (K) = \frac{1}{{}^*\Pi_T -K^2 - i \epsilon k_0} A^{\mu \nu}
    + \frac{1}{{}^*\Pi_L -K^2 - i \epsilon k_0} B^{\mu \nu}
    - \frac{\xi}{K^2 + i \epsilon k_0} D^{\mu \nu} ,
\end{equation}
where $^*\Pi_{L/T}$ is the HTL resummed longitudinal/transverse photon self-energy function [10].
$ A^{\mu \nu}$, $B^{\mu \nu}$ and $D^{\mu \nu}$ are the projection tensors [11], 
\begin{eqnarray}
 A^{\mu \nu} &=& g^{\mu \nu} - B^{\mu \nu}- D^{\mu \nu}, \\
 B^{\mu \nu} &=& - \tilde{K}^{\mu} \tilde{K}^{\nu}/K^2, \\
 D^{\mu \nu} &=& K^{\mu} K^{\nu}/K^2, 
\end{eqnarray}
where  $\tilde{K}=(k, k_0{\bf \hat{k}})$, $k=\sqrt{{\bf k}^2}$ 
and ${\bf \hat{k}}={\bf k}/k$ denotes the unit three vector along  ${\bf k}$.

 The parameter $\xi$ appearing in the term proportional to the projection tensor $D_{\mu \nu}$ represents the 
gauge-fixing parameter ($\xi=0$ in the Landau gauge). This gauge term plays an important role in the present analysis.

The vertex function is approximated by the tree (point) vertex. With the instantaneous exchange approximation
for the longitudinal gauge boson propagator, we get the DSEs for the three invariant functions $A(P)$, $B(P)$ and $C(P)$ 
\begin{eqnarray}
&  & -p^2[1-A(P)] = -e^2 \left. \int \frac{d^4K}{(2 \pi)^4}
       \right[ \{1+2n_B(p_0-k_0) \} \mbox{Im}[\ ^*G^{\rho \sigma}_R(P-K)]
       \times  \nonumber \\
  & & \Bigl[ \{ K_{\sigma}P_{\rho} + K_{\rho} P_{\sigma}
       - p_0 (K_{\sigma} g_{\rho 0} + K_{\rho} g_{\sigma 0} ) 
       - k_0 (P_{\sigma} g_{\rho 0} + P_{\rho} g_{\sigma 0} )
       + pkz g_{\sigma \rho} \nonumber \\
  & & + 2p_0k_0g_{\sigma 0}g_{\rho 0} \}\frac{A(K)}{[k_0+B(K)+i
       \epsilon]^2 - A(K)^2k^2 -C(K)^2 }
       + \{ P_{\sigma} g_{\rho 0} \nonumber \\
  & &  + P_{\rho} g_{\sigma 0} - 2p_0 g_{\sigma 0} g_{\rho 0} \}
       \frac{k_0+B(K)}{[k_0+B(K)+i \epsilon]^2 - A(K)^2k^2
       -C(K)^2 } \Bigr] 
       \nonumber \\ 
  & & + \{1-2n_F(k_0) \} \ ^*G^{\rho \sigma}_R(P-K) \mbox{Im} \Bigl[
       \{ K_{\sigma}P_{\rho}  + K_{\rho} P_{\sigma} - p_0 (K_{\sigma}
       g_{\rho 0} + K_{\rho} g_{\sigma 0} ) \nonumber \\
  & &  - k_0 (P_{\sigma}g_{\rho 0} + P_{\rho} g_{\sigma 0} ) + pkz g_{\sigma \rho} + 2p_0k_0g_{\sigma 0}g_{\rho 0}\}
       \times \nonumber \\
  & &  \frac{A(K)}{[k_0+B(K)+i \epsilon]^2 - A(K)^2k^2-C(K)^2 } 
       +  \{ P_{\sigma} g_{\rho 0} + P_{\rho} g_{\sigma 0} \nonumber \\
  & & \left. - 2p_0 g_{\sigma 0} g_{\rho 0} \}
       \frac{k_0+B(K)}{[k_0+B(K)+i \epsilon]^2 - A(K)^2k^2
       -C(K)^2 } \Bigr] \right] \ ,
\end{eqnarray}
\begin{eqnarray}
& & - B(P)= -e^2 \left. \int \frac{d^4K}{(2 \pi)^4} \right[
        \{1+2n_B(p_0-k_0)\} \mbox{Im}[\ ^*G^{\rho \sigma}_R(P-K)] \times
         \nonumber \\
  & & \Bigl[ \{ K_{\sigma} g_{\rho 0} + K_{\rho} g_{\sigma 0}
       - 2k_0 g_{\sigma 0} g_{\rho 0} \}
       \frac{A(K)}{[k_0+B(K)+i \epsilon]^2 - A(K)^2k^2
       -C(K)^2 } \nonumber \\
  & & + \{ 2g_{\rho 0} 2g_{\sigma 0} - g_{\sigma \rho} \} 
       \frac{k_0+B(K)}{[k_0+B(K)+i \epsilon]^2 - A(K)^2k^2-C(K)^2 }
       \Bigr] \nonumber \\ 
  & & + \{1-2n_F(k_0) \} \ ^*G^{\rho \sigma}_R(P-K) \mbox{Im} \Bigl[ \frac{A(K)}{[k_0+B(K)+i
       \epsilon]^2 - A(K)^2k^2 -C(K)^2 } \nonumber \\
  & & \times \{ K_{\sigma} g_{\rho 0} + K_{\rho} g_{\sigma 0} - 2k_0 g_{\sigma 0} g_{\rho 0} \} \nonumber \\
  & &  \left. + \frac{k_0+B(K)}{[k_0+B(K)+
       i \epsilon]^2 - A(K)^2k^2-C(K)^2 }
       \{ 2g_{\rho 0} 2g_{\sigma 0} - g_{\sigma \rho} \} \Bigr] 
       \right] \ , \\
& &  C(P) = -e^2 \int \frac{d^4K}{(2 \pi)^4} g_{\sigma \rho} 
       \{1+2n_B(p_0-k_0) \} \mbox{Im}[\ ^*G^{\rho \sigma}_R(P-K)]
       \times \nonumber \\
  & & \Bigl[ \frac{C(K)}{[k_0+B(K)+i \epsilon]^2 - A(K)^2k^2
       -C(K)^2 } + \{1-2n_F(k_0) \} \times \nonumber \\ 
  & & \left. \ ^*G^{\rho \sigma}_R(P-K) \mbox{Im} \Bigl[
       \frac{C(K)}{[k_0+B(K)+i \epsilon]^2 - A(K)^2k^2
       -C(K)^2 } \Bigr] \right] \ .
\end{eqnarray}

Above DSEs may have several solutions, and we choose the ``true'' solution by evaluating the effective 
potential $V[S_R]$ for the fermion propagator function $S_R$, then finding the lowest energy solution.
\begin{eqnarray}
 V [ S_R ] \!\! & =& \!\! i \mbox{Tr} \left[ P\!\!\!\!/ S_R \right]  +  i \mbox{Tr} \ln  \left[  i S_R^{-1} \right]  \nonumber \\
       & & - \frac{e^2}{2} \int \frac{d^4K}{(2 \pi)^4} \int \frac{d^4P}{(2 \pi)^4}
             \frac12 \mbox{tr} \left[ \gamma_{\mu} S_R(K) \gamma_{\nu} S_R(P) D_C^{\mu \nu} (P-K) \right. \nonumber \\
       & & \ \ \ \ \ \ \left.
               +  \gamma_{\mu} S_C(K) \gamma_{\nu} S_R(P) D_R^{\mu \nu} (P-K)
               +  \gamma_{\mu} S_R(K) \gamma_{\nu} S_C(P) D_A^{\mu \nu} (P-K) \right] ,  
\end{eqnarray}
\vspace*{0.3cm}
\begin{flushleft}
{\bf 2-2. Procedure to get the ``gauge-invariant'' solution}
\end{flushleft}

The function $A(P)$ above is nothing but the inverse of the fermion wave function renormalization constant $Z_2$, 
thus must be unity in order to satisfy the Ward-Takahashi identity in the ladder DSE analysis, where the 
vertex function receives no renormalization effect, $Z_1 =1$.
 
We must solve the above DSEs and get the solution satisfying the Ward-Takahashi identity $Z_2 = Z_1 (=1)$, 
where $Z_2 = A(P)^{-1}$. The procedure to get the ``gauge invariant'' solution is as follows; \\
     (1) Assume the nonlinear gauge such that the gauge parameter $\xi$ being a function of the 
momentum $K = (k_0, k)$ carried by the gauge boson. We parametrize $\xi$ as
\begin{equation}
         \xi (k_0, k) = \sum \xi_{mn} H_m(k_0) L_n(k) , \ \ \  k=\sqrt{{\bf k}^2} ,  
\end{equation}
where $\xi_{mn}$ are unknown parameters to be determined. $H_m$ and $L_n$ can in general be any 
ortho-normal functions, and we here take the Hermite functions for $H_m$ and the Laguerre functions for $L_n$. \\
    (2) When solving the above DSEs iteratively, impose the condition $A(P) =1$ by constraint for the input-functions
at each step of the iteration. \\
    (3) Determine $\xi_{mn}$ so as to minimize $|A(P) -1|^2$ for the output-functions and find the solutions for
$B(P)$ and $C(P)$.

\vspace*{0.5cm}
\begin{flushleft}
{\large\bf 3. ``Gauge invariant'' solution consistent with the Ward-Takahashi identity}
\end{flushleft}

Here we present the results obtained by the momentum-dependent guge parameter $\xi$. Number of parameters
$\xi_{mn}$ to minimize $|A(P)-1|^2$ is $2 \times 3 \times 2 =12$ (i.e., $m=0 \sim 2$ and $n=0, 1$) in the case of
complex $\xi$, and $4 \times 3=12$ (i.e., $m=0 \sim 3$ and $n=0 \sim 2$) in the case of real $\xi$. All the 
quantities with the mass dimension are evaluated in the unit of $\Lambda$, the cut-off parameter introduced
as usual to regularize the DSEs.

 Before presenting the ``gauge invariant'' solution, we show in Fig.1 the result of the critical temperature 
analysis for several values of constant $\xi$ to get a rough image for the size of gauge dependence. 
As can be seen clearly
the critical temperature strongly depends on the gauge, but the order of the phase transition does not.
\begin{figure}
\begin{center}
\includegraphics[width=10cm]{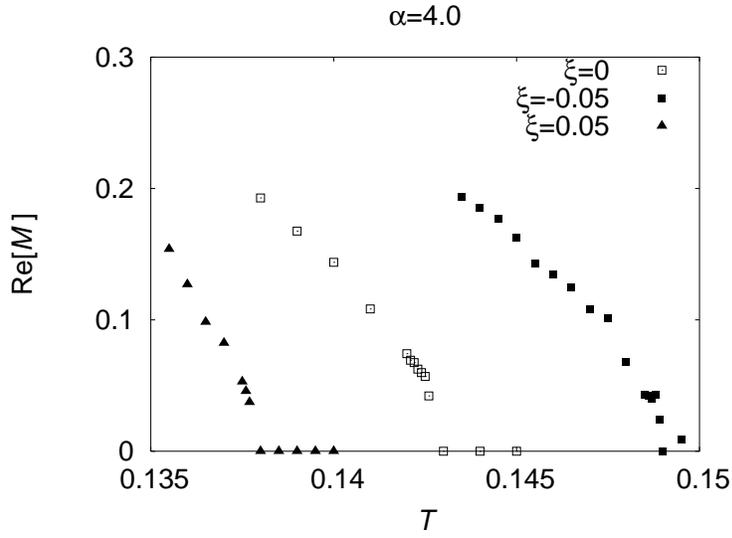} \\
\caption{\small Gauge-parameter-depnendence of the fermion mass Re[$M$] at the coupling constant $\alpha=4.0$ evaluated at
$p_0=0$, $p=0.1$.}
\label{aba:fig1}
\end{center}
\end{figure}

Now we present the solution consistent with the Ward-Takahashi identity, i.e., the ``gauge invariant'' solution. 
Analysis is now in progress, and the results shown below are, at present, still preliminary.

Firstly in Fig.2 we show $Re[A(P)]$. For comparison, results in the constant $\xi$ analyses are also shown in the
same figure. 
\begin{figure}
\begin{center}
\includegraphics[width=7cm,angle=270]{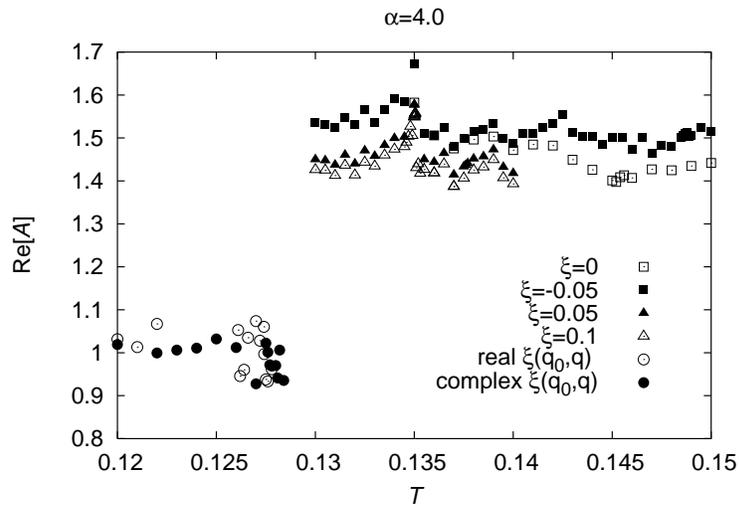}
\end{center}
\caption{\small Comparison of the wave function renormalization constant Re[$A$] at the coupling constant $\alpha=4.0$ 
evaluated at $p_0=0$, $p=0.1$.}
\label{aba:fig2}
\end{figure}

 Next let us study the property of the phase transition. Fig.3 shows the real part of the fermion mass $Re[M(P)]$, 
$M(P) \equiv C(P)/A(P)$ ($= C(P)$, because $A(P)=1$), obtained from the ``gauge invariant'' solution, as a function
of the temperature $T$. The mass is evaluated at $p_0=0$, $p=0.1$, to be consistent with the standard prescription 
to define the mass in the static limit, $p_0=0$, $p \to 0$.

Analyses to determine the critical temperature $T_c$, the critical coupling $\alpha_c$, and two critical 
exponents $\nu$ and $\eta$ are now in progress.

\begin{figure}
\begin{center}
\includegraphics[width=7cm,angle=270]{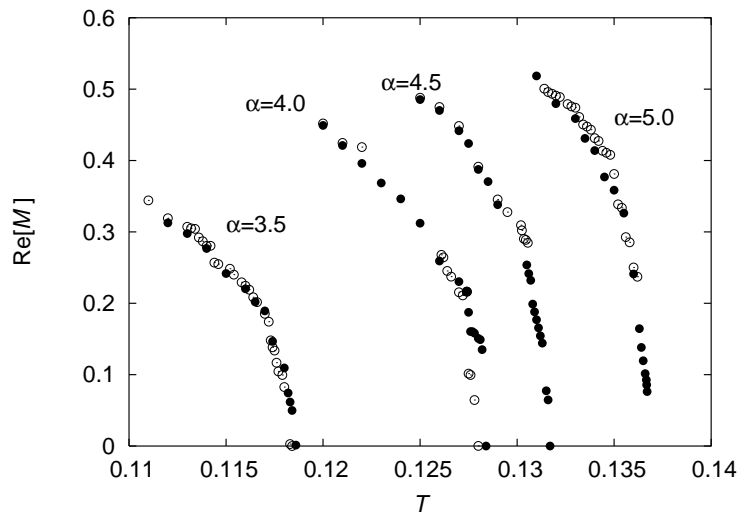} 
\end{center}
\caption{\small Temperature-dependence of the fermion mass Re[$M$] for various values of the coupling constant $\alpha=$ evaluated at
$p_0=0$, $p=0.1$, see text. Open circle denotes real $\xi$ data, while solid circle denotes complex $\xi$ data.}
\label{aba:fig3}
\end{figure}


    As can be clearly seen in the above Figures 2 and 3, two solutions obtained in the different two prescriptions, 
complex v.s. real gauge parameters $\xi$, show complete agreement. This fact indicates that the solution obtained
in the present procedure does not depend on the choice of the gauge parameter, namely that the solution is
``gauge invariant'' The results shown in this paper are still preliminary, and we will soon report the results of full
analysis.

The phase boundary curve in the $(T,\alpha)$-plane thus determined shows that the region of the symmetry broken phase
shrinks to the low-temperature and the strong-coupling side compared with that of the Landau gauge. This fact means that 
the effect of thermal fluctuation on the chiral symmetry breaking/restoration is smaller than that expected in the 
previous analysis in the Landau gauge [7]. 

\vspace*{0.5cm}
\begin{flushleft}
{\large\bf 4. Discussion and comments}
\end{flushleft}

Results presented in the present paper are still preliminary, because of the rough analysis of the data processing. 
We are now refining the data analysis and soon get the results of the thorough reanalysis. Though the main 
conclusion will not be altered, several important remarks should be added. \\
     (1) We performed the present analysis in two prescriptions for the nonlinear gauge parameter $\xi$, 
complex v.s. real. Allowing the gauge parameter $\xi$ to be a complex value may correspond to studying the 
non-hermite dynamics, thus may cause some troubles. In this sense we are interested in the result obtained by 
restricting the gauge parameter to the real value. What we found is a remarkable result: In both cases results
completely agree, thus getting a solution totally independent of the choice of gauges. This fact strongly 
indicates that we can get the gauge-invariant physical result by studying the DSE with the ladder interaction 
kernel through the present procedure. \\
     (2) In the present analysis, the consistency of the solution with the Ward-Takahashi identity is respected
 only by imposing the condition $A(P) \approx 1$.  Needless to say, in solving the (improved) ladder 
Dyson-Schwinger equation, there are no solutions totally consistent with the Ward-Takahashi identity. Despite this fact, 
following point should be closely examined: At least around or in the static limit, $p_0=0$, $p \to0$, where we 
calculated (defined) the mass, each invariant function $A(P)$, $B(P)$ or $C(P)$ should not have big momentum 
dependence. This condition may be important in connection with the consistency of the obtained solution with the
gauge invariance.  Result of the present analysis shows that at least $B(P)$ and $C(P)$ satisfy this condition. 

\vspace*{0.5cm}
\begin{flushleft}
{\large\bf References and footnotes}
\end{flushleft}
\begin{description}
\item{[1]} T. Maskawa and H. Nakajima, Prog. Theor. Phys. 52, 1326 (1974); 54, 860(1975);
       R. Fukuda and T. Kugo, Nucl. Phys. B117, 250 (1976).
\item{[2]} K. Yamawaki, M. Bando and K. Matumoto, Phys. Rev. Lett. 56, 1335 (1986);
       K.-I. Kondo, H. Mino and K. Yamawaki, Phys. Rev. D39, 2430 (1989).
\item{[3]} W. A. Bardeen, C. N. Leung and S. T. Love, Phys. Rev. Lett., 56, 1230 (1986);
       C. N. Leung, S. T. Love and W. A. Bardeen, Nucl. Phys. B273, 649 (1986);
       W. A. Bardeen, C. N. Leung and S. T. Love, Nucl. Phys. B323, 493 (1989).
\item{[4]} K.-I. Kondo and K. Yoshida, Int. J. Mod. Phys. A10, 199 (1995).
\item{[5]} M. Harada and A. Shibata, Phys. Rev. D59, 014010 (1998).
\item{[6]} K. Fukazawa, T. Inagaki, S. Mukaigawa and T. Muta, Prog. Theor. Phys. 105, 
       979 (2001).
\item{[7]} Y. Fueki, H. Nakkagawa, H. Yokota and K. Yoshida,  Prog. Theor. Phys. 110,  
       777 (2003); 
       H. Nakkagawa, H. Yokota, K. Yoshida and  Y. Fueki, Pramana -- J. of Phys. 60,  
       1029 (2003);
       Y. Fueki, H. Nakkagawa, H. Yokota and K. Yoshida, Prog. Theor. Phys. 107, 759
       (2002).
\item{[8]} A. Rebhan, Phys. Rev. D46, 4779 (1992).
\item{[9]} H. Nakkagawa, H. Yokota and K. Yoshida, hep-ph/0703134, to appear in the proceedings of the International
       Workshop on ``The Origin of Mass and Strong Coupling Gauge Theories (SCGT06)'', Nov. 22-24, 2006, Nagoya
       University, Nagoya, Japan.
\item{[10]} V. V. Klimov, Sov. J. Nucl.Phys. 33, 934 (1981); Sov. Phys. JETP 55, 199 (1982);
        H. A. Weldon, Phys. Rev. D26, 1394 (1982); Phys. Rev. D26, 2789 (1982). 
\item{[11]} H. A. Weldon, Ann. Phys. (N.Y.) 271, 141 (1999). 

\end{description}
\end{document}